\documentclass{jfm}

\usepackage{graphicx}
\usepackage{newtxtext}
\usepackage{newtxmath}
\usepackage{natbib}
\usepackage{hyperref}
\usepackage{subcaption}
\usepackage{caption} % justified fig captions
\hypersetup{
    colorlinks = true,
    urlcolor   = blue,
    citecolor  = black,
}
\usepackage[capitalize]{cleveref}
\usepackage{bm}

\newcommand{\imagi}{\mathrm{i}}
\newcommand{\dd}{\mathrm{d}}

\newcommand{\RomanNumeralCaps}[1]
\title{An ensemble of Gaussian fields with multifractal statistics for turbulence}

\author{Mark Warnecke\aff{1}
  \corresp{\email{mwarneck@uci.edu}},
  Lukas Bentkamp\aff{2}, Gabriel B. Apolin\'ario\aff{2}, Michael Wilczek \aff{2}, \and Perry Johnson\aff{1} 
 }

\affiliation{\aff{1}Mechanical \& Aerospace Engineering, University of California Irvine.
\aff{2}Theoretical Physics I, University of Bayreuth, Universitätsstraße 30, 95447 Bayreuth, Germany.}

\begin{document}
\maketitle

\begin{abstract}
Obtaining accurate field statistics continues to be one of the major challenges in turbulence theory and modeling. From the various existing modeling approaches, multifractal models have been successful in capturing intermittency in velocity gradient and increment distributions. On the other hand, superstatistical models from nonequilibrium statistical mechanics have shown the capacity to model PDFs of various statistical turbulent quantities as ensembles of simpler stochastic processes. Here, we present an approach that generates field statistics in the form of a characteristic functional by promoting a model for multifractal increment statistics to an ensemble of Gaussian fields. By carefully designing the correlation function and the corresponding weight of each subensemble, we are able to define a functional that exhibits multifractal two-point inertial-range and dissipation-range statistics, and that blends into realistic large-scale behavior. Additionally, the method is capable of producing multifractal statistics with any of the widely used singularity spectra. We characterize the fidelity of our approach through comparisons to literature results from direct numerical simulations. Overall, our framework thereby bridges between three different perspectives to turbulence: superstatistics, multifractals, and functional approaches to turbulence.
\end{abstract}

% \begin{keywords}
% Authors should not enter keywords on the manuscript, as these must be chosen by the author during the online submission process and will then be added during the typesetting process (see \href{https://www.cambridge.org/core/journals/journal-of-fluid-mechanics/information/list-of-keywords}{Keyword PDF} for the full list).  Other classifications will be added at the same time
% \end{keywords}

% {\bf MSC Codes }  \inlinetodo{{\it(Optional)} Please enter your MSC Codes here}

\section{Introduction}\label{sec:1_intro}

Turbulence is characterized by velocity fluctuations whose statistics are strongly scale-dependent. Probability density functions (PDFs) of the velocity increments are approximately Gaussian at separation distances commensurate to and larger than the integral length scale, but display heavier tails as the separation length scale decreases through the inertial range and into the dissipative range of scales. This phenomenon, known as intermittency, presents a major challenge to creating accurate turbulence models. Over the years, a number of techniques have been developed to model intermittency, such as multifractals and superstatistics, which are reviewed in the following.

In the language of multifractal theory, multiscale observables can be written as a statistical mixture of a continuum of H\"{o}lder exponents. Each exponent $h$ occupies a set of dimension $D(h)$ and this determines the contribution of each $h$ to the full statistics \citep{mandelbrot1977fractals,parisi1985multifractal,falconer1990fractals,meneveau1991multifractal}. Among the observables to which multifractal theory has been successfully applied are velocity gradients \citep{Nelkin1990prA, Chevillard2006lagrangegrad, Ishihara2009DNSgrad, Johnson2017gradient, PEREIRA2018grad, Luo2022grad}, Eulerian velocity increments \citep{benzi1984multifractal, meneveau1991multifractal}, Lagrangian velocity increments \citep{biferale2004lagrangian,arneodo2008lagrangian,Chevillard2005lagrange,chevillard2012crp}, and Lagrangian acceleration \citep{biferale2004lagrangian}. Crucially, it provides a connection between the asymptotic scaling of all of these observables in terms of a single $D(h)$ function. %

At first glance, superstatistics \citep{beck2003superstats,beck2004cmt,beck2011superstats} represents a complementary approach to model intermittency based on a mixture distribution. Applied to nonequilibrium statistical mechanics, superstatistics describes a mixture of Boltzmann statistics with the temperature as the mixing parameter \citep{beck2003superstats,beck2004cmt,beck2011superstats}. In the context of turbulence, the mixing parameter may be, for instance, the energy dissipation.
This description has been shown to capture non-Gaussian turbulence statistics both in the Eulerian \citep{beck2000eulersuperstats} and Lagrangian frames \citep{reynolds2003lagrangiansuperstats,reynolds2005lagrangiansuperstats,beck2003lagrangian,beck2007prl}.
It is also worth noting that the multifractal model can be formulated in a way akin to superstatistics. Specifically, \cite{chevillard2012crp} considered a superposition of Gaussian increment statistics with a fluctuating variance. This indicates that intimate relations between the two approaches exist.

Previous applications of the multifractal and superstatistical approaches have focused on the modeling of  individual random variables, e.g., two-point statistics. While the multifractal language provides a connection between the universal properties of different observables \citep{benzi1991gradients,benzi1998multiscale,chevillard2012crp}, the interest in extending these approaches to modeling the statistics of entire random fields remains. \cite{wilczek2016njp}
showed that a superposition of Gaussian characteristic functionals (i.e., a statistical ensemble of Gaussian fields) with a fluctuating correlation function is capable of producing non-Gaussianity and intermittency. 
This naturally leads to the question whether a mixture model with realistic multifractal properties comprising entire field statistics can be developed. A first step into this direction was recently taken by \cite{friedrich2021jpc}, who developed a model for field statistics with classic K62 multifractal scaling~\citep{kolmogorov1962refinement, obukhov1962some}. The approach has been extended to enable a statistical refinement of sparse data sets by \cite{lubke2023jpc} and has furthermore been applied to the stochastic modeling of turbulent wind fields \citep{friedrich2022prxen}.
We mention as well a complementary approach to field statistics, where realizations of a random field are generated with properties compatible with multifractal theory, either in a static \citep{chevillard2019skewed} or dynamic way \citep{apolinario2022dynamical,beck2024numerical}.

In this paper, we establish a relation between existing multifractal models and the superstatistical Gaussian field ensemble approach.
We construct an ensemble of Gaussian fields that features increment and gradient statistics complying with a given multifractal singularity spectrum, albeit only for its even moments. From a conceptual point of view this highlights close connections between the discussed previous approaches and lays the foundation for a future connection to statistical field theories based on the Navier-Stokes equation such as the functional approach \citep{hopf1952jrma, ohkitani2020study}.

The content is arranged as follows.
In \cref{sec:2_background}, we review the  multifractal modeling of \citet{chevillard2012crp} and the ensemble of Gaussian fields of \citet{wilczek2016njp}.
In \cref{sec:3_general_transform}, we show how the multifractal description can be generalized to a superstatistical characteristic functional. \cref{sec:4_lognormal} demonstrates the approach with a simple example---the classic lognormal model. In \cref{sec:5_general}, a more general model is considered, with an arbitrary singularity spectrum, and realistic large-scale and viscous-range behavior. Conclusions are drawn in Section \ref{sec:6_conclusions}.

\section{Theoretical background}\label{sec:2_background}

\subsection{Gaussian mixtures for multifractal velocity increment PDFs}
\label{sec:multifractal_increments}
In this section, we summarize the formulation of multifractals as Gaussian mixtures put forward by \cite{chevillard2012crp}. 
% 
% \\
% 
% 
% 
% 
% 
% 
%    
% 
The paper considers velocity increments $v = \delta_\ell u$ across a scale $\ell$ in the inertial range, i.e.,~$\eta \ll \ell \ll L$ where $\eta$ denotes the Kolmogorov scale and $L$ the integral scale of the flow. The velocity increment, conditional on a given H\"older exponent $h$, is modeled as a zero-mean Gaussian random variable, %
with PDF
\begin{equation}
  g( v;S(\ell;h) ) = \frac{1}{\sqrt{2 \pi S(\ell;h)}} \exp\left[ -\frac{v^2}{2 S(\ell;h)} \right] \, .
\end{equation}
Notice that we have adopted a notation for the Gaussian PDF $g(x; \sigma^2)$ where $\sigma^2$ is the variance. Here, $S(\ell;h)$ is the second-order structure function conditional on a given H\"older exponent, and behaves as
\begin{equation} \label{eq:chevillard_second_order_sf}
S(\ell;h) = 2 \sigma_u^2 \left( 1 + \left(\frac{L}{\ell}\right)^2\right)^{-h} \, ,
\end{equation}
in the inertial and large-scale ranges. Compared to~\cite{chevillard2012crp}, we have added a large-scale saturation. The large-scale root mean square velocity is denoted $\sigma_u$, and $S(\ell,h)$ correctly asymptotes to $2 \sigma_u^2$ as $\ell \to \infty$. Additionally, this function shows an $(\ell/L)^{2h}$ scaling in the inertial range.
In this framework, the PDF of velocity increments is given by 
\begin{equation} \label{eq:incr_pdf_h_superposition}
    f_v(v ; \ell) = \int \dd h \, f_{h}(h;\ell) \, g(v;S(\ell;h)) \, ,
\end{equation}
where
$f_{h}(h;\ell)$ is the probability density of H\"older exponents at fixed scale $\ell$.
This mixture of Gaussian random variables with fluctuating variance naturally gives rise to non-Gaussian statistics.

The mixture distribution $f_h(h;\ell)$ provided by multifractal phenomenology, including the inertial and large-scale ranges, is given by
\begin{equation} \label{eq:h_pdf_general}
    f_h(h; \ell) = \begin{cases}
        \frac{1}{Z(\ell)} \,
	\left( 1+ (L/\ell)^2 \right)^{-(1-D(h))/2},&\text{if }h_{\min} \leq h \leq h_{\max}\,, \\
  0, &\text{otherwise,}\end{cases}
\end{equation}
where $D(h)$ is the multifractal singularity spectrum, $h_{\min}$ and $h_{\max}$ are bounds on the H\"older exponent (explained below), and $Z(\ell)$ is the normalization of the PDF.
A key assumption of the multifractal model, relevant for the subsequent discussion, is that $D(h)$ is assumed to be independent of $\ell$ over the inertial range of scales.

With this formulation, the velocity increment displays multifractal scaling exponents, at least for its even moments,
\begin{equation} \label{eq:strucfunc_exponents}
    \langle v^n \rangle \sim \ell^{\zeta_n}, \
    \text{where } \
    \zeta_n = \min_h \left[ nh + 1 - D(h) \right] ,
\end{equation}
as can be shown with a steepest descent approximation in the limit of vanishing scale separation~\citep{chevillard2012crp}. Due to the use of Gaussians, all odd-order moments of the velocity increments are zero. We remark, however, that \cite{chevillard2012crp} 
addresses the skewness of the velocity increments by modifying the Gaussian base distribution. 

There are several empirical models for $D(h)$ in the turbulence literature, and in \cref{sec:3_general_transform} we consider a change of variables where $D(h)$ can be freely specified.
As explicit examples, we consider three representative models of singularity spectra: The lognormal, with a quadratic spectrum
\begin{equation}
    D(h) = 1 - \frac{(h-h_0)^2}{2 \sigma_h^2}
    \label{eq:quadratic_d_of_h};
\end{equation}
the She-L\'ev\^eque model \citep{SheLeveque1994spectrum}, with spectrum
\begin{equation}
    D(h)
    = -1 + 3\left[ \frac{1+\log\log\left(\frac{3}{2}\right)}{\log\left(\frac{3}{2}\right)} - 1 \right]\left(h - \frac{1}{9}\right) - \frac{3}{\log\left(\frac{3}{2}\right)}\left(h - \frac{1}{9}\right)\log\left(h-\frac{1}{9}\right);
\end{equation}
and the Sreenivasan-Yakhot model \citep{SreenivasanYakhot2021moments}, defined by
\begin{equation}
    D(h) = 1 - c - \beta h + 2 \sqrt{c \beta h}.
\end{equation}
The free parameters of the lognormal model can be determined by fits to the statistics of direct numerical simulation (DNS) or experimental data. Typical values are $h_0 = 0.37$, $\sigma_h^2 = 0.025$ \citep{chevillard2012crp}. For the Sreenivasan-Yakhot model, $\beta = 3c - 3$ and $c = 7.3$ are predicted by \citep{SreenivasanYakhot2021moments}.

The range of the H\"older exponents, $h_{\min} \leq h \leq h_{\max}$, important for a complete description of the multifractal exponents, requires accurate statistical data at high Reynolds numbers to be reliably inferred. This topic has recently received attention, with numerical simulations done at increasing spatial resolution \citep{buaria2023role} and novel techniques for measuring exponents at negative order \citep{lashermes2008comprehensive}.
Estimates of $h_{\min}$ range from $-0.88$ \citep{debue2018dissipation,dubrulle2019beyond} to $0.15$ \citep{lashermes2008comprehensive}.
For this study, we rely on simple estimates by choosing $h_{\min}$ and $h_{\max}$ as the solutions to $D(h)=0$.

In the remainder of this paper we introduce a change of variables that allows to map the multifractal framework as outlined above to an ensemble of Gaussian fields. 

\subsection{Ensemble of Gaussian characteristic functionals for non-Gaussian field statistics}\label{sec:1_intro_gaussian_ensemble}

While the statistics of velocity increments (and two-point statistics, more generally) have been a central focus for turbulence theory over the past century, they provide an incomplete characterization of turbulent fields.
A complete description is provided, for instance, by the characteristic functional, satisfying a closed and infinite-dimensional dynamical equation (the Hopf equation, see \cite{hopf1952jrma,monin1971vol1and2}). This section reviews the approach of \cite{wilczek2016njp}, where non-Gaussian and intermittent characteristic functionals are built as mixtures (ensembles) of Gaussian random fields (compare also~\cite{bentkamp2019natcom,friedrich2021jpc,friedrich2022prxen,lubke2023jpc}).
While not able to describe skewness, this approach provides a simple and analytically tractable basis for capturing the statistics of turbulent fields. %

In the following we consider one-dimensional fields for simplicity. When comparing theoretical results to data from experiments or DNS data, we interpret these fields \textit{cum grano salis} as one-dimensional cuts through three-dimensional fields. An extension of the approach to three-dimensional vector fields is briefly outlined in Appendix \ref{app:vector_fields}.
For a field $u(x)$, the two-point statistics are encapsulated in the following joint PDF,
\begin{equation}
    f_2(u_1, u_2) = \left\langle \delta(u_1 - u(x_1)) \delta(u_2 - u(x_2))\right\rangle,
\end{equation}
or alternatively in its Fourier transform, the two-point characteristic function,
\begin{equation}
    \phi_2(\alpha_1, \alpha_2) = \left\langle \exp \left[ \imagi \left( \alpha_1 u(x_1) + \alpha_2 u(x_2) \right) \right] \right\rangle.
\end{equation}
Field statistics include the $N$-point statistics for any number of points $N$. This can be captured in a single object, the characteristic functional, given by the expression \citep{kolmogorov1935transformation,hopf1952jrma}
\begin{equation}
    \label{eq:characteristic_functional}
    \Phi[\alpha] = \left\langle \exp\left[ \imagi \int_{-\infty}^{\infty} \dd x \, \alpha(x) u(x) \right] \right\rangle.
\end{equation}
Here, the argument of $\Phi$ is the test function $\alpha(x)$. This characteristic functional contains complete information of the probability densities of observing any velocity field configuration $u(x)$.

In particular, the characteristic functional of a zero-mean Gaussian field is
\begin{equation}
    \label{eq:gaussian_characteristic_functional}
    \Phi^{G}[\alpha; \gamma] = \exp \left[ -\frac{1}{2}\int_{-\infty}^{\infty} \dd x\int_{-\infty}^{\infty} \dd x' \alpha(x) \widetilde C(x-x';\gamma) \alpha(x') \right],
\end{equation}
where $\widetilde C(x-x';\gamma)=\langle u_{\gamma}(x)u_{\gamma}(x')\rangle $ denotes the two-point velocity covariance function. We have added $\gamma$ as a parametric dependence. We restrict ourselves to statistically homogeneous fields, such that the covariance function is an even function depending only on $x - x'$. 
Note that the two-point covariance is sufficient to determine the entire field statistics of the Gaussian field.
This provides a practical advantage in developing theories based on Gaussian field statistics, and non-Gaussianity may be recovered by superposing the statistics of an ensemble of Gaussian fields \citep{wilczek2016njp}. In particular, the covariance function can be chosen to vary depending on a parameter $\gamma$, allowing for its interpretation as a conditional covariance function. In this way, the Gaussian fields become subensembles of the full model. The characteristic functional for the resulting ensemble statistics thus takes the form
\begin{equation}
    \Phi[\alpha] = \int \dd \gamma \, f_{\gamma}(\gamma) \, \Phi^G[\alpha;\gamma]
    \label{eq:Gaussian_fields_ensemble}
\end{equation}
where $f_{\gamma}(\gamma)$ is the PDF of the parameter $\gamma$. This parameter could be related, for example, to a fluctuating length scale, the local dissipation rate or, as discussed below, the fluctuating H\"{o}lder exponent of a turbulent field. Note that, by restricting ourselves to zero-mean Gaussian fields, we do not capture odd-order moments. As a result, our model statistics lacks skewness. The possibility of addressing skewness within our framework, in analogy to the discussion in~\cite{chevillard2012crp}, is left for future investigation.

The apparent similarity between \cref{eq:incr_pdf_h_superposition} for expressing multifractal two-point statistics \citep{chevillard2012crp} and \cref{eq:Gaussian_fields_ensemble} for the superstatistics of random fields invites a closer inspection of the relationship between these two approaches.
This is done in the next section, and for that we outline the statistics of velocity increments and gradients that we obtain from the functional description. The velocity increment statistics are obtained by choosing the test function \citep{wilczek2016njp, monin1971vol1and2}
\begin{equation} \label{eq:test_function_increment}
  \alpha(x) = \Delta \left[ \delta(x - y - \ell)- \delta(x - y) \right],
\end{equation}
which, inserted into \cref{eq:Gaussian_fields_ensemble}, yields the characteristic function of the increment at scale $\ell$
\begin{equation} \label{eq:incr_charf_gamma_superposition}
  \phi_v(\Delta;\ell) = \int \dd \gamma \, f_{\gamma}(\gamma) \, \exp\left( -\frac{1}{2} \widetilde S(\ell;\gamma) \Delta^2 \right),
\end{equation}
with the conditional structure function
\begin{equation} \label{eq:strucfunc_corr_relation}
  \widetilde S(\ell;\gamma) = 2 \left[ \widetilde C (0;\gamma) - \widetilde C (\ell;\gamma) \right].
\end{equation}  
The Fourier transform of this characteristic function produces the increment PDF
\begin{equation} \label{eq:incr_pdf_gamma_superposition}
    f_{v}(v;\ell) = \int \dd \gamma \, f_{\gamma}(\gamma) \, g\left(v;\widetilde S(\ell; \gamma)\right) .
\end{equation}
Furthermore, the gradient statistics can be found in a similar way, by choosing the test function \citep{wilczek2016njp}
\begin{equation}
  \label{eq:test_function_gradient}
  \alpha(x) = -\beta \frac{\partial}{\partial x} \delta(x - y).
\end{equation}
Using integration by parts before evaluating the delta function, one obtains the characteristic function of the gradient $A = \partial_x u$
\begin{equation}
  \phi_A(\beta) = \int \dd \gamma \, f_{\gamma}(\gamma) \, \exp\left( \frac{1}{2} \widetilde C''(0;\gamma) \beta^2 \right) = \int \dd \gamma \, f_{\gamma}(\gamma) \, \exp\left( - \frac{1}{4} \widetilde S''(0;\gamma) \beta^2 \right),
\end{equation}
where primes denote derivatives with respect to $\ell$. A Fourier transform returns the PDF of the velocity gradient,
\begin{equation}
    \label{eq:general_grad_pdf}
    f_{A}(A) = \int \dd \gamma f_{\gamma}(\gamma) \, g\left(A;\frac{1}{2}\widetilde S''(0; \gamma)\right) \, .
\end{equation}

\section{Connecting multifractal mixtures and Gaussian ensembles}
\label{sec:3_general_transform}

We may now build a connection between the multifractal formulation of \cite{chevillard2012crp} and the ensemble of Gaussian fields of \cite{wilczek2016njp}. Let us start with the multifractal formulation, in particular \cref{eq:incr_pdf_h_superposition} which we repeat:
\begin{equation} \label{eq:incr_pdf_h_repeat}
  f_v(v;\ell) = \int \dd h \, f_h(h;\ell) \, g(v;S(\ell; h)).
\end{equation}
For this distribution, $h$ is the mixing parameter. 
On the other hand, notice that \cref{eq:incr_pdf_gamma_superposition}, obtained from the ensemble of Gaussians, is also a mixture, this time with the scale-independent mixing distribution $f_{\gamma}$:
\begin{equation} \label{eq:incr_pdf_gamma_repeat}   
  f_v(v;\ell) = \int \dd \gamma \, f_\gamma(\gamma) \, g\left(v;\widetilde S(\ell; \gamma)\right).
\end{equation}

In \cref{eq:incr_pdf_h_repeat}, the scale dependence of $f_h$ is a requirement of multifractal phenomenology, while in \cref{eq:incr_pdf_gamma_repeat}, its absence in $f_{\gamma}$ is necessary for the functional formulation: As can seen in \cref{eq:Gaussian_fields_ensemble}, the mixing distribution $f_\gamma(\gamma)$ cannot depend on the scale of the observable, which only comes in through the test function \cref{eq:test_function_increment} and thus becomes an argument of the subensemble structure function $\widetilde{S}$.
In order to write an ensemble of Gaussian fields that is compatible with the multifractal formulation of \cref{eq:incr_pdf_h_repeat}, we introduce a change of variables which connects $\gamma$ and $h$.

Let us introduce the cumulative distribution functions (CDFs)
\begin{align}
F_h(h;\ell) &= \int_{-\infty}^h \mathrm{d}h' \, f_h(h';\ell), \ \text{and} \\
F_\gamma(\gamma) &= \int_{-\infty}^\gamma \mathrm{d}\gamma' \, f_\gamma(\gamma') \, .
\end{align}

We claim that, given the PDFs $f_\gamma(\gamma)$ and $f_h(h;\ell)$, as well as a structure function $S(\ell; h)$, both superpositions are equivalent, i.e.,
\begin{equation}
 f_v(v;\ell) = \int\mathrm{d}\gamma \, f_\gamma(\gamma) \, g\left(v; \widetilde S(\ell; \gamma) \right) = \int\mathrm{d}h \, f_h(h;\ell) \, g(v;S(\ell; h)), 
\end{equation}
given that
\begin{equation}
	\widetilde S(\ell; \gamma) = S\left(\ell; F_h^{-1}(F_\gamma(\gamma);\ell) \right) \, .\label{eq:s_tilde_general_transform}
\end{equation}

To show this, we choose the following $\ell$-dependent substitution that maps $\gamma$ to $h$:
\begin{equation} \label{eq:h_general_transform}
  h(\gamma;\ell) = F_h^{-1}(F_\gamma(\gamma);\ell),
\end{equation}
such that
\begin{equation} \label{eq:h_gamma_cdfs}
  F_h(h(\gamma;\ell);\ell) = F_\gamma(\gamma) \, .
\end{equation}
A derivative with respect to $\gamma$ results in
\begin{equation}
\frac{\mathrm d}{\mathrm d \gamma} F_h(h(\gamma;\ell);\ell) = \frac{\mathrm d F_h(h(\gamma;\ell);\ell)}{\mathrm d h} \frac{\mathrm d h}{\mathrm d \gamma} = f_h(h;\ell) \frac{\mathrm d h}{\mathrm d \gamma} = \frac{\mathrm d F_\gamma(\gamma)}{\mathrm d \gamma} = f_\gamma(\gamma),
\end{equation}
and thus the usual change of probability measures
\begin{equation}
f_h(h;\ell) \, \mathrm dh = f_\gamma(\gamma) \, \mathrm d\gamma \, .
\end{equation}
Applying this substitution to the integral, we can write
\begin{align}
\int\mathrm{d}\gamma \, f_\gamma(\gamma) \, g\left(v; \widetilde S(\ell; \gamma) \right) &= \int\mathrm{d}\gamma \, f_\gamma(\gamma) \, g\left(v; S\left(\ell; F_h^{-1}(F_\gamma(\gamma);\ell) \right) \right) \\
&= \int\mathrm{d}h \, f_h(h;\ell) \, g(v;S(\ell; h)),
\end{align}
which concludes the argument.
This means that the increment statistics from an $h$ ensemble with any scale-dependent PDF $f_h$ can be reproduced by a $\gamma$ ensemble with a scale-independent PDF $f_{\gamma}$, subject to an appropriate change of the structure function. Note that only the transformed structure function $\widetilde S$ is fixed by the transform, whereas the parameter distribution $f_\gamma(\gamma)$ can be chosen freely in advance.

Therefore, we can transform a multifractal model for velocity increment PDFs with arbitrary singularity spectrum into a Gaussian mixture increment PDF with scale-independent mixing distribution. The latter expression can then be generalized to a mixture (ensemble) of characteristic functionals. The resulting functional reproduces the increment statistics of the original multifractal model. Furthermore, it is capable of producing multifractal velocity gradient statistics as well. In \citet{chevillard2012crp}, in comparison, velocity gradient statistics are constructed through a compatibility argument between random variables at different scales.
We illustrate the change of variables \cref{eq:h_general_transform} with various examples in the following sections.

\section{Example: Lognormal model}
\label{sec:4_lognormal}
As a first example to illustrate the mapping from the multifractal formulation to the ensemble of Gaussian fields, we choose a quadratic singularity spectrum, $D(h) = 1 - \frac{(h-h_0)^2}{2\sigma_h^2}$ (Eq.~\ref{eq:quadratic_d_of_h}), the lognormal multifractal model, without bounds on $h$ ($h_{\min}=-\infty$, $h_{\max}=\infty$). 
According to \cref{eq:h_pdf_general}, the resulting PDF $f_h$ is Gaussian,
\begin{equation}
  f_h(h;\ell) = \frac{1}{Z(\ell)}\exp\left[ - \log\left( 1 + \left(\frac{L}{\ell}\right)^2 \right) \frac{(h-h_0)^2}{4\sigma_h^2} \right] = g\left(h - h_0; \Sigma^2(\ell)\right)\,,
\end{equation}
where $Z(\ell)$ is the appropriate normalization, and
\begin{align}
    \Sigma^2(\ell) &= \frac{2\sigma_h^2}{\log\left( 1 + \left(\frac{L}{\ell}\right)^2 \right)}
\end{align}
is the $\ell$-dependent variance.

Its CDF reads
\begin{equation}
  F_h(h;\ell) = \int_{-\infty}^h \, \mathrm{d}h' f_h(h';\ell) = G\left(\frac{h-h_0}{\Sigma(\ell)}\right)\,,
\end{equation}
where $G(x)$ is the CDF of a standard Gaussian. As laid out in the previous section, we want to express the increment PDFs given by \cref{eq:incr_pdf_h_repeat} as a superposition with a scale-independent mixing parameter $\gamma$ (\cref{eq:incr_pdf_gamma_repeat}). Since we can freely choose the distribution of $\gamma$, we set it to be a standard Gaussian, with PDF $f_\gamma(\gamma) = g(\gamma;1)$ and CDF $F_\gamma(\gamma) = G(\gamma)$. The mapping is given by \cref{eq:h_general_transform}:
\begin{align}
    h(\gamma; \ell) &= F_h^{-1}(F_\gamma(\gamma); \ell)
    = h_0 + \Sigma(\ell) G^{-1}(G(\gamma))
    = h_0 + \Sigma(\ell) \gamma \,.
\end{align}
For the superposition, \cref{eq:incr_pdf_gamma_repeat}, we also need the structure function conditional on $\gamma$, $\widetilde S$.
To this end, we choose the $h$-dependent structure function $S(\ell,h)$ according to the multifractal formulation \cref{eq:chevillard_second_order_sf},
from which we compute the $\gamma$-dependent structure function by~\cref{eq:s_tilde_general_transform}:
\begin{equation} \label{eq:strucfunc_transformed}
\begin{split}
    \widetilde S(\ell; \gamma) &= S(\ell; h(\gamma; \ell)) \\
    &= 2\sigma_u^2 \left( 1 + \left(\frac{L}{\ell}\right)^2\right)^{-h(\gamma; \ell)} \\
    &= 2\sigma_u^2 \left( 1 + \left(\frac{L}{\ell}\right)^2\right)^{-h_0 - \gamma\sqrt{\frac{2\sigma_h^2}{\log(1+(L/\ell)^2)}}}\,.
\end{split}
\end{equation}
Through~\cref{eq:strucfunc_corr_relation}, this translates into the $\gamma$-dependent covariance function
\begin{equation}
    \widetilde C(\ell; \gamma) = \sigma_u^2 - \sigma_u^2 \left( 1 + \left(\frac{L}{\ell}\right)^2\right)^{-h_0 - \gamma\sqrt{\frac{2\sigma_h^2}{\log(1+(L/\ell)^2)}}}\,.
\end{equation}
which can be used to construct the full ensemble of Gaussian fields, \cref{eq:Gaussian_fields_ensemble}. Therefore, we have defined a fully explicit characteristic functional, comprehensively describing field statistics, including lognormal multifractal increment statistics.

While analytically simple, the model allows for arbitrarily negative H\"older exponents, which are unphysical. This issue can be fixed by adding bounds $h_{\min} \leq h \leq h_{\max}$, as discussed in \cref{sec:multifractal_increments}. The corresponding PDF of H\"older exponents, \cref{eq:h_pdf_general}, then becomes a truncated Gaussian,\begin{equation}
  f_h(h;\ell) = \begin{cases} \frac{1}{Z(\ell)}\exp\left[ - \log\left( 1 + \left(\frac{L}{\ell}\right)^2 \right) \frac{(h-h_0)^2}{4\sigma_h^2} \right], &\text{if }h_{\min} \leq h \leq h_{\max}\,, \\
  0, &\text{otherwise.}\end{cases}
\end{equation}
where $Z(\ell)$ is a new appropriate normalization. Its CDF is
\begin{equation}
  F_h(h;\ell) = \int_{-\infty}^h \, \mathrm{d}h' f_h(h';\ell) = \frac{G\left(\frac{h-h_0}{\Sigma(\ell)}\right) - G\left(\frac{h_{\min}-h_0}{\Sigma(\ell)}\right)}{G\left(\frac{h_{\max}-h_0}{\Sigma(\ell)}\right) - G\left(\frac{h_{\min}-h_0}{\Sigma(\ell)}\right)}\,, \quad \text{if }h_{\min} \leq h \leq h_{\max}\,,
\end{equation}
with which we compute a new mapping,
\begin{align}
    h(\gamma; \ell) &= F_h^{-1}(F_\gamma(\gamma); \ell) \\
    &= h_0 + \Sigma(\ell) \, G^{-1}\left(G\left(\frac{h_{\min}-h_0}{\Sigma(\ell)}\right) + G(\gamma) \left(G\left(\frac{h_{\max}-h_0}{\Sigma(\ell)}\right) - G\left(\frac{h_{\min}-h_0}{\Sigma(\ell)}\right)\right)\right)\,.
\end{align}
This expression can be used in \cref{eq:strucfunc_transformed} for an improved version of the Gaussian ensemble with lognormal statistics.

Recently, a Gaussian ensemble with lognormal statistics was introduced by \cite{friedrich2021jpc}. This approach relies on the combination of a fractional Ornstein-Uhlenbeck process with a rescaling of space depending on a lognormal, scale-independent random parameter. It can be shown that the specific transformation in \cite{friedrich2021jpc} to construct the ensemble of characteristic functionals corresponds to the lognormal case of the general transform method presented here, but with the scale-dependent parameter being the coarse-grained dissipation rate $\varepsilon_\ell$ instead of the H\"older exponent $h$.
More details are provided in \cref{app:Friedrich}.

\section{General multifractal Gaussian ensemble} \label{sec:5_general}
The mapping \cref{eq:h_general_transform} is applicable to general H\"older exponent PDFs $f_h(h;\ell)$ and structure functions $S(\ell;h)$.  In this section, we apply this mapping to a multifractal model with an arbitrary spectrum of exponents. We include bounds for the H\"older exponent, as in the previous section, and add a realistic viscous-range cut-off to the formulation. The main goal of this section is to demonstrate the capacity of the model to reproduce the features of general multifractal models for the increment and gradient statistics of turbulence. For the sake of presentation, we make particular choices for the form of $f_h$, $S$, and various parameters, to be able to evaluate the model numerically and compare to DNS data.

In general, various modeling choices for $S(\ell;h)$ and $f_h(h ;\ell)$ are possible as long as they have certain asymptotic scalings in the dissipation, inertial, and large-scale ranges. The onset of the dissipation range is built into the multifractal formalism by local Kolmogorov scales $\eta(h) \ll L$ that are functions of the H\"older exponent. They depend on the Reynolds number in the following way \citep{paladin1987degrees,Nelkin1990prA,frisch1995turbulence}:
\begin{equation}
  \eta(h) = L R^{-\tfrac{1}{1+h}} \, ,
\end{equation}
where $R = \Rey/\Rey_*$, with $\Rey$ the large-scale Reynolds number and $\Rey_*$ a tuning coefficient. A general H\"older exponent PDF $f_h(h;\ell)$, for $h_{\min} \leq h \leq h_{\max}$, must have the asymptotic scaling:
\begin{equation} \label{eq:f_multifractal_scaling}
    f_h(h;\ell) \sim \begin{cases}
    
    {\displaystyle \left( \frac{\eta(h)}{L}\right)^{1-D(h)}\,,}&\text{if }\ell \ll \eta(h)\,, 

    \\{\displaystyle \left(\frac{\ell}{L} \right)^{1-D(h)}\,,}&\text{if } \eta(h) \ll \ell \ll L\,,
    
    \\{1\,,}&\text{if } L \ll \ell \,.
    
    \end{cases}
\end{equation}
Additionally, a general $S(\ell;h)$ must have the asymptotic scaling:
\begin{equation} \label{eq:S_multifractal_scaling}
    \frac{S(\ell;h)}{2 \sigma_u^2} \sim \begin{cases}
    
    {\displaystyle \left(\frac{\ell}{\eta(h)} \right)^{2} \left(\frac{\eta(h)}{L} \right)^{2h}  \,,}&\text{if }\ell \ll \eta(h)\,, 

    \\{\displaystyle \left(\frac{\ell}{L} \right)^{2h}\,,}&\text{if } \eta(h) \ll \ell \ll L\,,

    \\1&\text{if } L \ll \ell \,.
    
    \end{cases}
\end{equation}
Using the general transformation outlined in~\cref{sec:3_general_transform}, any such choice of $f_h$ and $S$ can be used to construct a Gaussian ensemble with multifractal scaling. In the dissipative range, if the conditional structure function, $S(\ell;h)$,
matches \cref{eq:S_multifractal_scaling}, then it shows smooth behavior for a fixed H\"older exponent.   
Therefore, the gradient variance of each subensemble is then readily computed as
\begin{equation} \label{eq:grad-var-subensemble}
  \sigma_A^2 = \lim_{\ell \rightarrow 0} \frac{S(\ell;h)}{\ell^2} = 2 \frac{\sigma_u^2}{L^2} R^{  \frac{2(1-h)}{1+h}}  \, .
\end{equation}

For the subsequent discussion, let us specify particular forms of $f_h$, $S$, and the distribution of $\gamma$. For the conditional structure function, we choose the following extension to \cref{eq:chevillard_second_order_sf}:
\begin{equation} \label{eq:multi_scale_strucfunc_s}
		S(\ell;h) = 2 \sigma_u^2 \frac{\left(1+(L/\ell)^2\right)^{-h}}{(1+(\eta(h)/\ell)^{2 \lambda} )^{(1-h)/\lambda}} \, ,
\end{equation}
which is essentially how dissipation range statistics are included in~\citet{chevillard2012crp}.    
The additional parameter $\lambda$ is a tuning coefficient that controls the onset of the dissipation range.
Analogous to \cref{eq:multi_scale_strucfunc_s}, we also add a viscous-range cut-off to the distribution of H\"older exponents:
\begin{equation} \label{eq:multi_scale_general_h_pdf}
	f_h(h;\ell) = \begin{cases}{\displaystyle \frac{1}{Z(\ell)} \, \frac{\left( 1+ (\eta(h)/\ell)^{2 \lambda} \right)^{(1-D(h))/2\lambda}}{\left( 1+ (L/\ell)^2 \right)^{(1-D(h))/2}}\,,}&\text{if }h_{\min} \leq h \leq h_{\max}\,, \\
  0, &\text{otherwise,}\end{cases}
\end{equation}
where $Z(\ell)$ is the normalization of this PDF. The unfixed model parameters $\Rey_*$ and $\lambda$ are chosen in \cref{sec:inertial_range} by matching our model with the DNS results in Fig. \ref{fig:increment_pdf} for the case $n = 2$.

Since the distribution of the scale-independent parameter $\gamma$ can be freely chosen, for simplicity, we opt here for a uniform distribution between $0$ and $1$. As a result, the ensemble of functionals~\cref{eq:Gaussian_fields_ensemble} can be written as
\begin{equation}
    \label{eq:general_functional}
    \Phi [\alpha] = \int_0^1 \dd \gamma\, \Phi_G[\alpha; \gamma]\,.
\end{equation}
Recognizing that the CDF of $\gamma$ is $F_\gamma(\gamma) = \gamma$ (for $\gamma$ between $0$ and $1$),~\cref{eq:h_gamma_cdfs} tells us that $\gamma$ is equal to the CDF of the H\"{o}lder exponent,
\begin{equation}
\label{eq:beta_transform}
\gamma(h; \ell) = F_h(h;\ell) = \int_{-\infty}^h \dd h' f_h(h' ;\ell).
\end{equation}

\subsection{Visualizing the transform}
\begin{figure}
    \centering
    \includegraphics[width=1.0\linewidth]{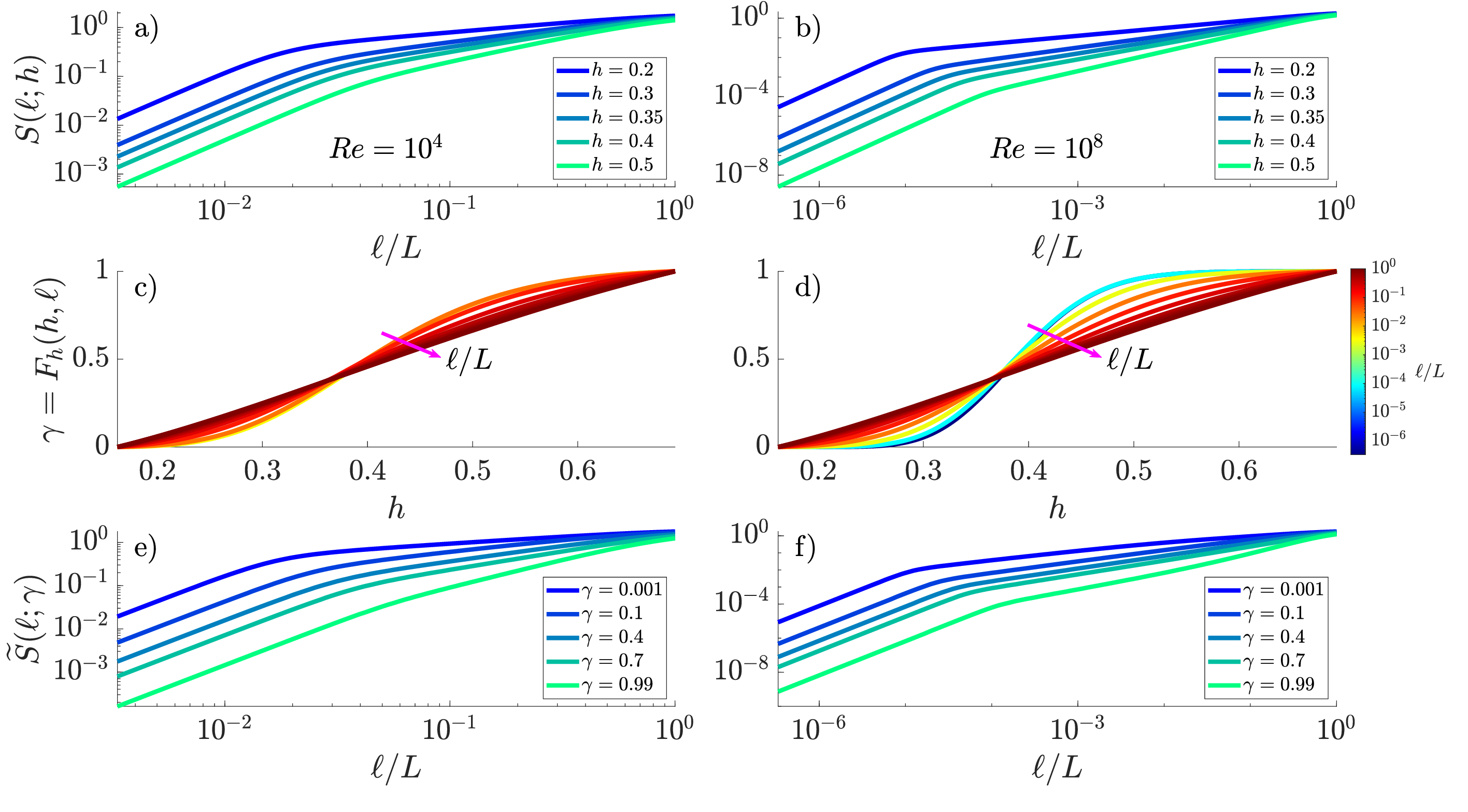}
    \caption{a), b): The second-order structure functions at fixed H\"{o}lder exponents for $\Rey = 10^4$ (left) and $\Rey = 10^8$ (right). c), d): The scale-independent parameter $\gamma$ plotted versus the H\"older exponent with a magenta arrow indicating the direction of small $\ell$ to large $\ell$. e), f): The second-order structure function of selected Gaussian subensembles corresponding to fixed values of $\gamma$. For this figure, the H\"{o}lder exponent defining the transformation has a She-L\'ev\^eque singularity spectrum and varies from 0.162 to 0.694.}
    \label{fig:structure_function_and_transform}
\end{figure}

This transform, \cref{eq:beta_transform}, is used to map from the multifractal to the functional formulation, as outlined in~\cref{sec:3_general_transform}. \cref{fig:structure_function_and_transform} illustrates its behavior, by showing the structure functions $S(\ell;h)$ (top row), the scale-dependent transform between $h$ and $\gamma$ (middle row), and the transformed structure functions $\widetilde{S}(\ell; \gamma)$ (bottom row), for the case of the She-L\'ev\^eque singularity spectrum.
Each column corresponds to a different Reynolds number. The transform (middle row) is given by the CDF of the H\"{o}lder exponents \eqref{eq:beta_transform}, which depends on the length scale $\ell$. At large scales, the CDF is flatter, corresponding to a broad distribution of $h$. Due to the large-scale cut-off, the CDFs ultimately converge to a single curve (dark red). At small scales, the CDFs are steeper, corresponding to a narrower $h$-distribution, also converging to a single curve due to the viscous cut-off (dark blue). Each value of $\gamma$ corresponds to a single subensemble in our ensemble of Gaussian fields, but maps to a range of H\"{o}lder exponents seen by the different curves for each length scale. At one point in the center, corresponding to the most likely value of $h$, the dependence of the transform on the length is weak. For $\gamma$-values further away from this center, the corresponding $h$-values change with length scale, as visualized by the differently colored curves.

The upper and lower rows show how the $h$-dependent structure functions are transformed into the $\gamma$-dependent structure functions according to~\cref{eq:s_tilde_general_transform}. This transformation is computed by numerically inverting~\cref{eq:beta_transform}. For the larger Reynolds number, the $h$-dependent structure functions (b) show clear inertial-range power-law scaling, built in~\cref{eq:multi_scale_strucfunc_s}. The transformed structure functions (f) are still strongly reminiscent of this, even though their exponents change with length scale. (compare with \cref{eq:strucfunc_transformed}).

\begin{figure}
    \centering
    \includegraphics[width=1.0\linewidth]{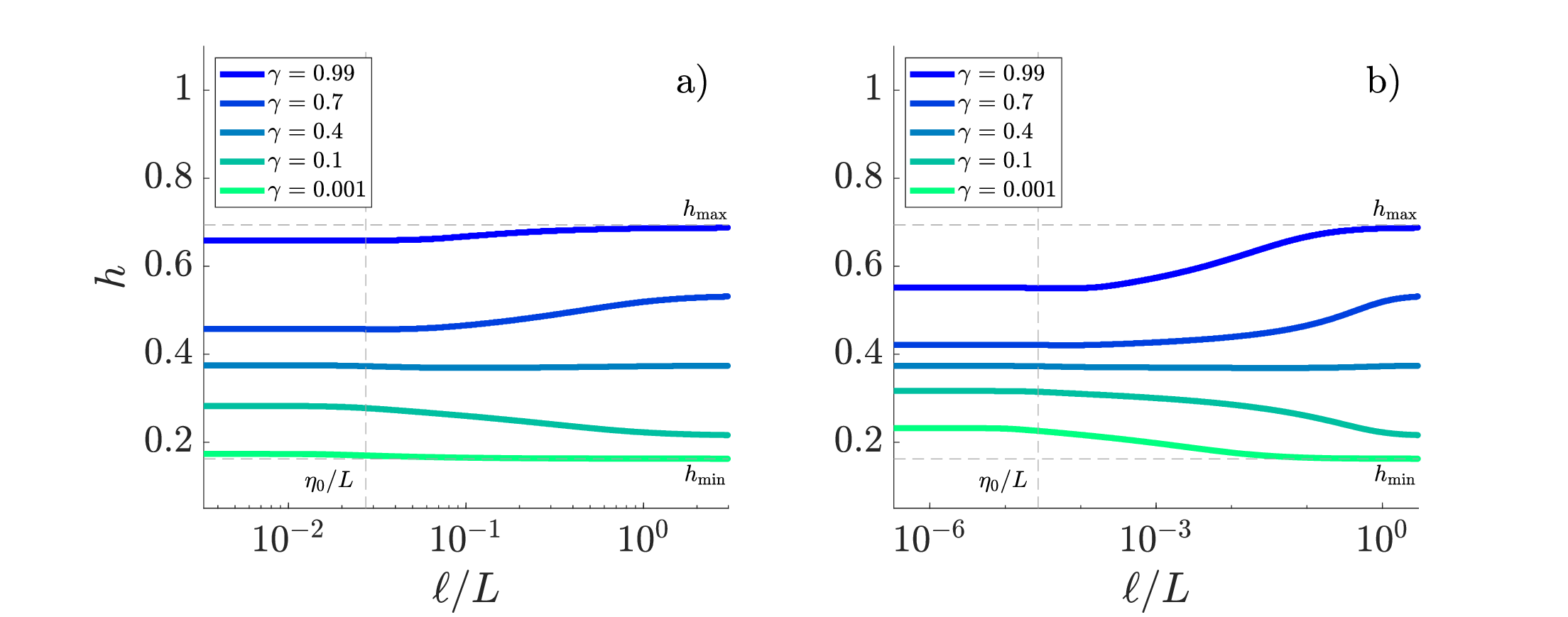}
    \caption{The variation of the H\"{o}lder exponent with separation distance of selected members of the ensemble are plotted for Reynolds numbers equal to $10^4$ and $10^8$. We define $\eta_0 = \eta(h=1/3)$ to illustrate the onset of the dissipation range. For this figure, the H\"{o}lder exponent defining the transformation has a She-L\'ev\^eque singularity spectrum and varies from 0.162 to 0.694.}
    \label{fig:h_variation}
\end{figure}

% 

% 

%
%.
%
%

Another way of looking at our transform is to consider the different $h$-values that a single $\gamma$-value maps to across length scales. This is shown in~\cref{fig:h_variation} for two different Reynolds numbers. While $\gamma$-values around $0.4$ correspond to almost constant $h$-values, other $\gamma$-curves cover a wider range of $h$-values. It is this length dependence in the transform that allows us to encode multifractal statistics using a single distribution for $\gamma$ (in this case, the uniform distribution between $0$ and $1$). \cref{fig:h_variation} shows that the variation with $\ell$ becomes more pronounced as the Reynolds number increases, since the transform has to accommodate very wide (at large scales) and very narrow (at small scales) $h$-distributions.

\subsection{Inertial-range statistics}\label{sec:inertial_range}
\begin{figure}
    \centering
    \includegraphics[width=1.0\linewidth]{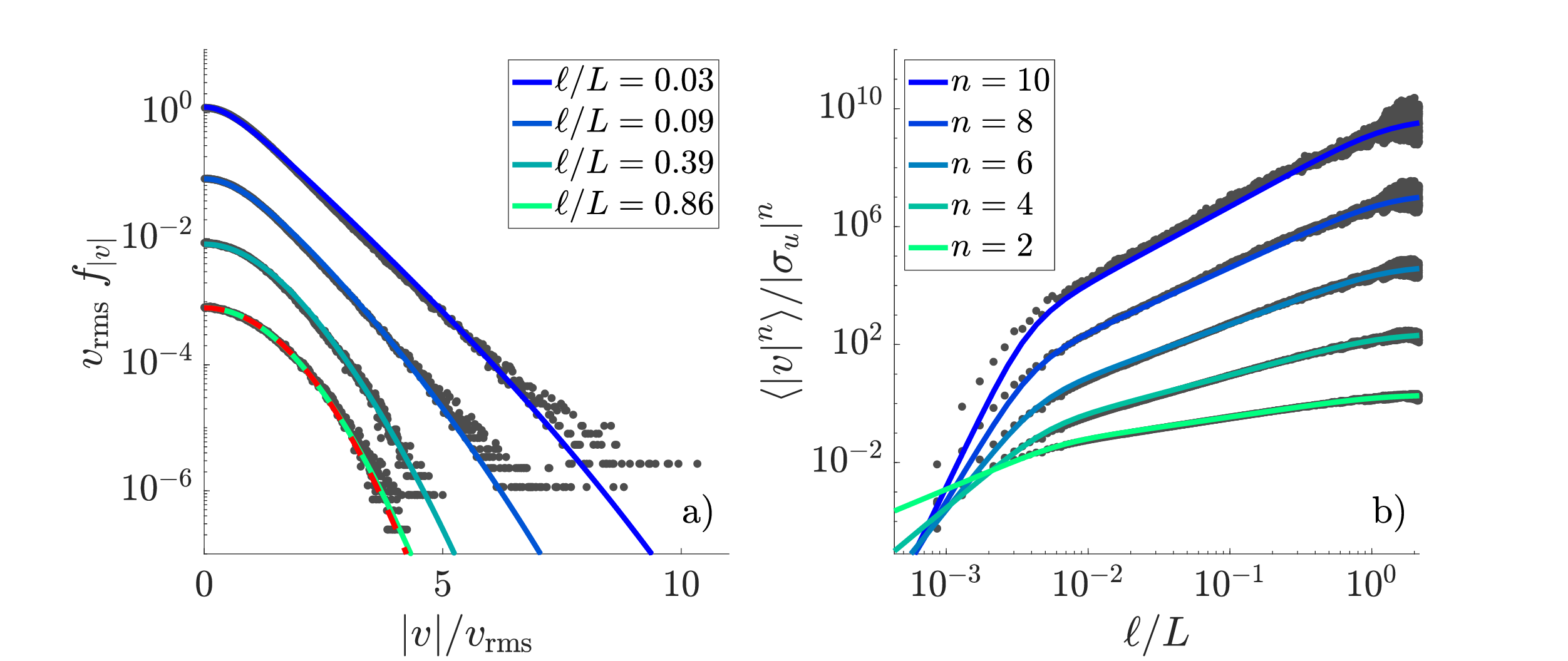}
    \caption{(a) The standardized velocity increment PDF at various separation distances from the Gaussian ensemble (lines) plotted against simulation data from the JHTDB~\citep{JHUTurbulence} \texttt{iso8192} dataset (points). Additionally a standardized Gaussian is plotted in a red dashed line for reference.  (b) The $n$-th order structure function vs.\ separation length ($\ell$) from the Gaussian ensemble (lines) plotted against simulation data from the JHTDB~\citep{JHUTurbulence, YeungTurbulence} \texttt{iso8192} dataset (points). For this figure, the H\"{o}lder exponent defining the transformation has a She-L\'ev\^eque singularity spectrum and varies from 0.162 to 0.694. All curves and points are vertically shifted for clarity. (\cite{JHUTurbulence}, \cite{YeungTurbulence}).}
    \label{fig:increment_pdf}
\end{figure}

To illustrate how our ensemble of Gaussian fields exhibits multifractal features resembling the statistics of turbulence, we compute increment statistics by numerically evaluating \cref{eq:incr_pdf_gamma_superposition}. 
Longitudinal increment PDFs and structure functions from the model are compared to DNS data in~\cref{fig:increment_pdf} (we used the \texttt{iso8192} dataset from the Johns Hopkins Turbulence Database (JHTDB), see \citet{JHUTurbulence}).
\cref{fig:increment_pdf}a shows that the model reproduces quantitatively the transition from close-to-Gaussian statistics at large scales to the heavy-tailed distributions characterizing the small scales.
The corresponding structure functions are shown in~\cref{fig:increment_pdf}b. Note that these are the unconditional structure functions predicted by the model, not to be confused with the $h$- and $\gamma$-dependent second-order functions $S(\ell; h)$ and $\widetilde{S}(\ell; \gamma)$ used to construct the model. Model parameters in \cref{eq:multi_scale_strucfunc_s,eq:multi_scale_general_h_pdf} were chosen to visually fit the $\langle |v|^2 \rangle/|\sigma_u|^2$ curve in~\cref{fig:increment_pdf}b: Their values are $\lambda = 2$, $\Rey_* = 82$ and $ L = 0.67 L_D$, where $L_D$ is the integral length scale as reported in the \texttt{iso8192} dataset. In~\cref{fig:increment_pdf}b we observe clear power law scaling in the inertial range matching the DNS data, as well as the proper asymptotic behavior in the viscous and large scale cut-offs.

\begin{figure}
    \centering
    \includegraphics[width=1\linewidth]{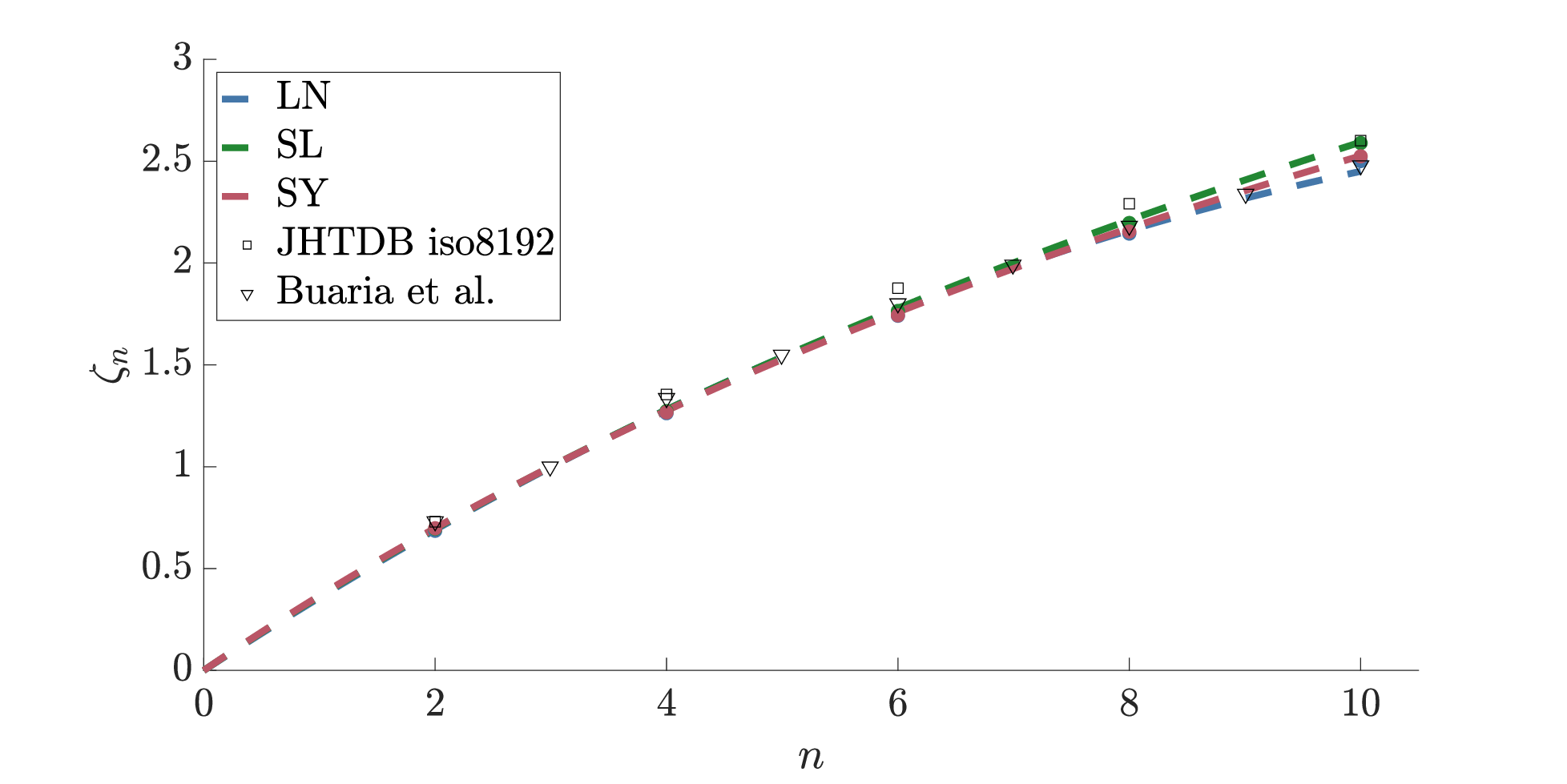}
    \caption{Scaling exponents of the even $n$th-order structure functions for various singularity spectra---Lognormal (LN), She-L\'ev\^eque (SL, \cite{SheLeveque1994spectrum}), and Sreenivasan-Yakhot (SY, \cite{SreenivasanYakhot2021moments})---calculated from the Gaussian ensemble (colored points) and compared with numerical calculations of Eq. (\ref{eq:strucfunc_exponents}), obtained from multifractal theory
    (colored dashed lines). The minimum and maximum H\"{o}lder exponents are set at the points where $D(h) =0$. Additionally simulation data is plotted from the JHTDB \texttt{iso8192} dataset (\cite{JHUTurbulence}, \cite{YeungTurbulence}) and \cite{buaria2023saturation}. }
    \label{fig:strucfunc_scaling}
\end{figure}

By multiplying~\cref{eq:incr_pdf_gamma_superposition} with powers of the increments and integrating analytically, we get
\begin{equation}\label{eq:velocity_difference_moments}
    \langle v^{n} \rangle = (n-1)!! \int_0^1 \dd \gamma \, \widetilde{S}(\ell; \gamma)^{n/2}\,, 
\end{equation}
for even $n$, which allows us to compute structure functions of any order directly from the second-order subensemble structure functions $\widetilde{S}(\ell; \gamma)$. 
With our change of variables, \cref{eq:h_general_transform}, one can show that the corresponding scaling exponents agree with \cref{eq:strucfunc_exponents}.
In order to check this numerically, we measure the scaling of the structure functions in \cref{eq:velocity_difference_moments} in the limit $\Rey \rightarrow \infty$. To this end, we choose an interval of length scales much smaller than $L$. Then the best linear fit to a double logarithmic plot of \cref{eq:velocity_difference_moments} is used to compute $\zeta_n$. In~\cref{fig:strucfunc_scaling}, we compare the exponents computed from the model's structure functions (data points) to the exponents expected by the Legendre transform \cref{eq:strucfunc_exponents} (dashed lines), which agree reasonably well. This confirms that the ensemble of Gaussian fields constructed here is capable of reproducing the velocity increment scaling of various multifractal models.

\subsection{Gradient statistics}
\begin{figure}
    \centering
    \includegraphics[width=1.0\linewidth]{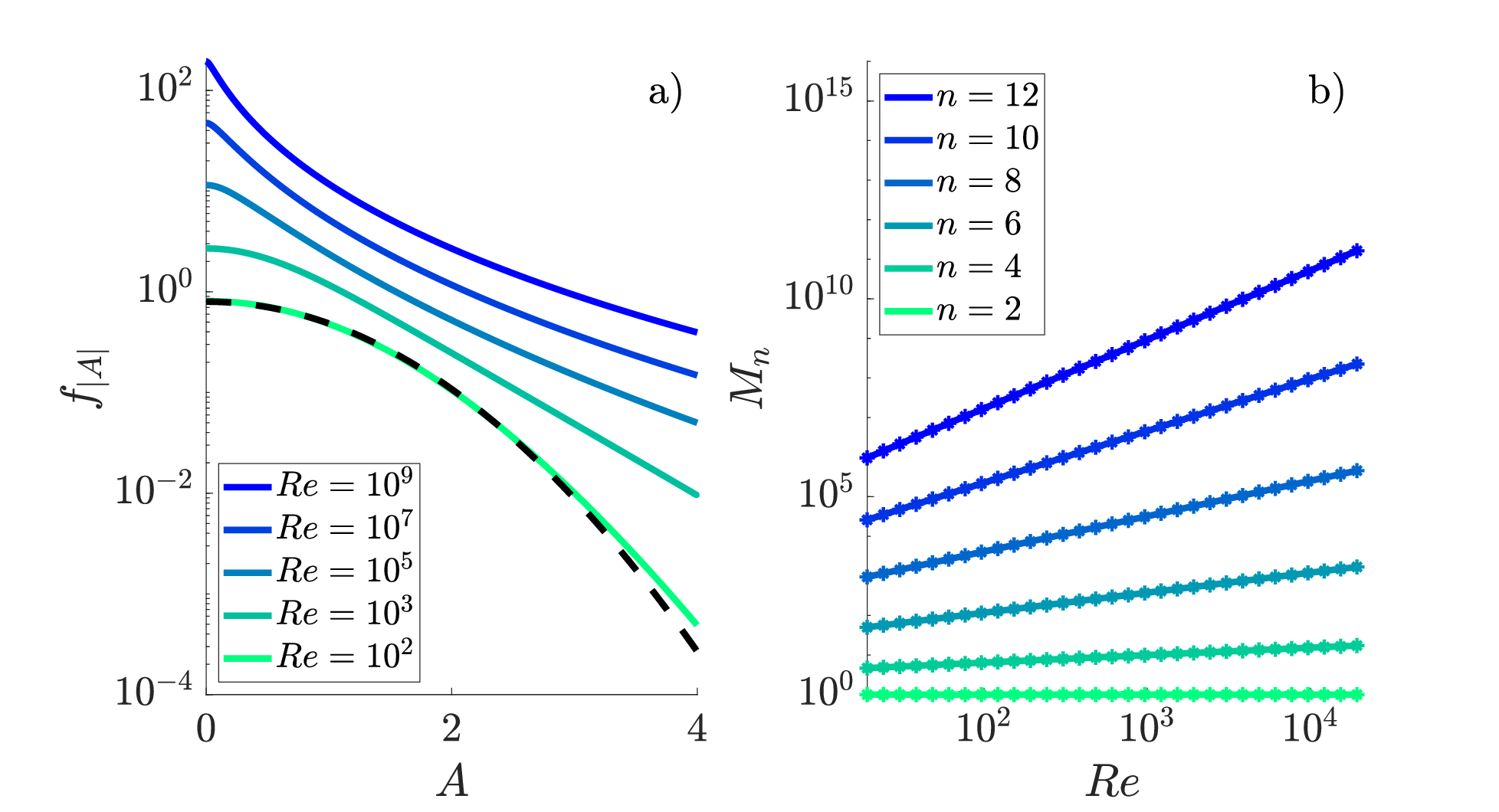}
    \caption{(a) Standardized gradient PDFs from the Gaussian ensemble (solid lines), alongside a standardized Gaussian (dashed line) for comparison. (b) The $n$th-order standardized gradient moment from the Gaussian ensemble based on superposition of $\gamma$ (Eq. \eqref{eq:grad_moments}, lines) plotted on top of calculations from the corresponding version of Eq. \eqref{eq:grad_moments} based on superposition of H\"{o}lder exponents (points) versus Reynolds number.  The H\"{o}lder exponents defining the transformation have a She-L\'ev\^eque singularity spectrum and vary from 0.162 to 0.694. The plots are vertically shifted for clarity.}
    \label{fig:grad_pdf}
\end{figure}

Based on the full multi-scale description we introduced at the beginning of~\cref{sec:5_general}, our model is able to predict gradient statistics as well. The tuning parameters ($\lambda$, $\Rey_*$, and $L$) are set by matching the second-order structure function to data (see~\cref{sec:inertial_range}). The gradient PDF, obtained from the functional with the right choice of the test function (\cref{eq:test_function_gradient}), is given by a superposition of Gaussian PDFs,~\cref{eq:general_grad_pdf}. The PDF of the absolute value of the gradient is shown at different Reynolds numbers in \cref{fig:grad_pdf}a, standardized by its variance, the average of \cref{eq:grad-var-subensemble}.

\begin{figure}
    \centering
    \includegraphics[width=1.0\linewidth]{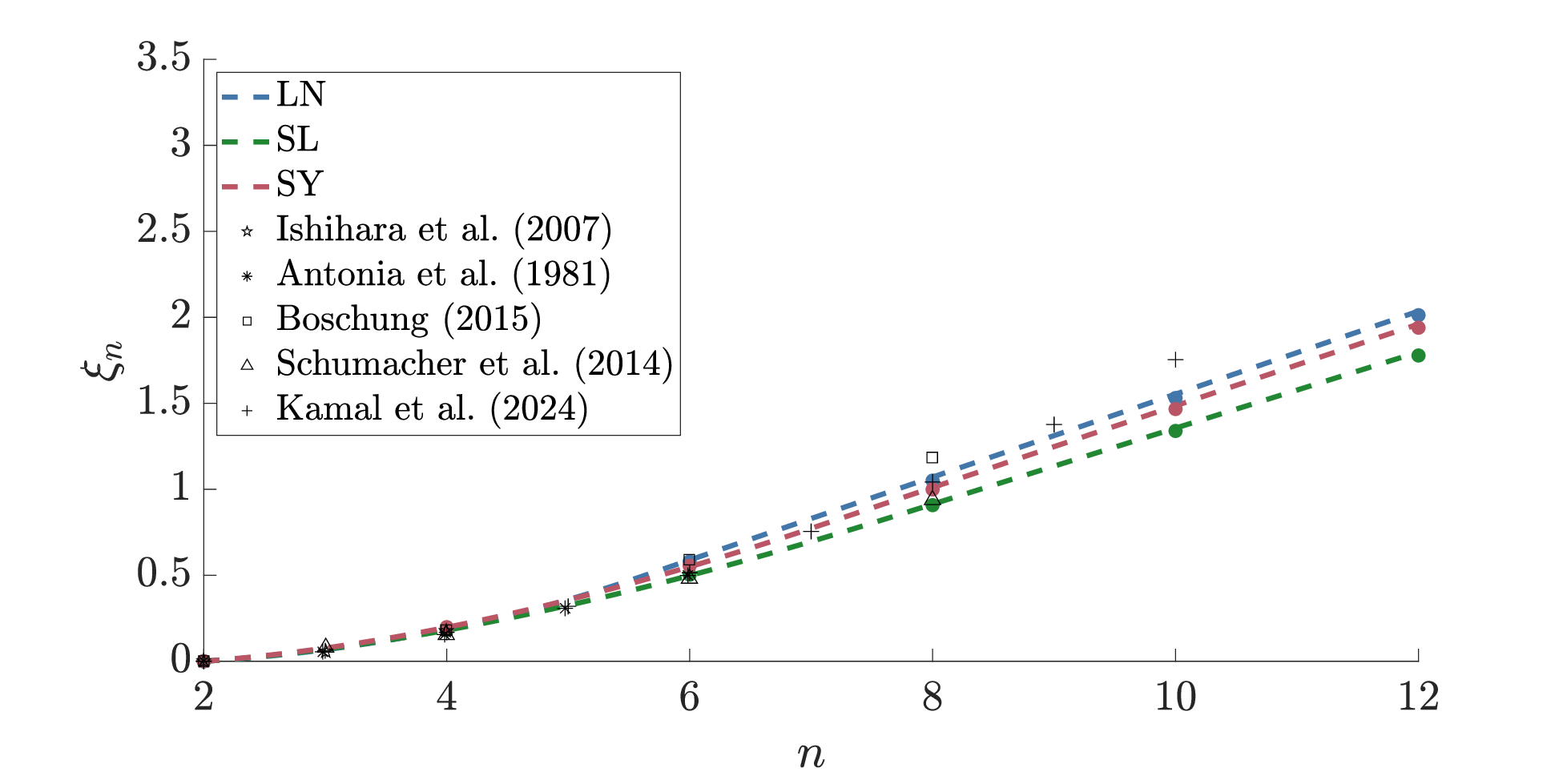}
    \caption{Reynolds number scalings of the standardized velocity gradient moments calculated from the Gaussian ensemble (colored points) and from \cref{eq:grad_scaling_exponents} (dashed curves) for the Lognormal (LN), She-L\'ev\^eque (SL), and Sreenivasan-Yakhot (SY) singularity spectra. The minimum and maximum H\"{o}lder exponents are set at the points where $D(h) =0$. DNS and experimental results ($\square$: \cite{boschung2015graddatasquare}, $\bigtriangleup$: \cite{schumacher2014graddatauptriangle}, $\bigtriangledown$: \cite{ishihara2007graddatadowntriangle}, $\triangleright$: \cite{arneodo1981graddatarighttriangle} , and $\triangleleft$: \cite{kamal2024data})}
    \label{fig:grad_exponents}
\end{figure}

Analogous to the increment moments, gradient moments can be computed by multiplying~\cref{eq:general_grad_pdf} with powers of the gradient and integrating analytically. As a result, gradient moments can be written as
\begin{equation} \label{eq:grad_moments}
    \langle  A^n \rangle = (n-1)!!\int_0^1 \dd \gamma \, \left( \frac{1}{2} \widetilde{S}''(0;\gamma) \right)^{n/2}.
\end{equation}
Again, this allows us to compute moments directly from the subensemble structure functions $\widetilde{S}(\ell; \gamma)$. We show the standardized moments,
\begin{equation}
    M_n = \frac{\langle A^n \rangle }{\langle A^2 \rangle^{n/2}}\,,
\end{equation}
as a function of Reynolds number in~\cref{fig:grad_pdf}b and compare them to a corresponding calculation made with an ensemble of H\"{o}lder exponents to demonstrate the validity of the model. They are expected to scale as a power of the Reynolds number:
\begin{equation}
    M_n \sim \Rey^{\xi_n}\,.
\end{equation}
To check their scaling, we compute the exponents as linear best fits to the double logarithmic plot in \cref{fig:grad_pdf}b. The result is shown in~\cref{fig:grad_exponents}, and compared to data from a few different DNS and to the exponents obtained from multifractal theory~\citep{Nelkin1990prA},
\begin{equation}
    \label{eq:grad_scaling_exponents}
    \rho_n \approx - \min_h  \frac{n(h-1)+1-D(h)}{1+h},
    \hspace{0.2\linewidth}
    \xi_n = \rho_n - \frac{n}{2}\rho_2 \,.
\end{equation}
The comparison with the expected exponents in~\cref{fig:grad_exponents} shows that they agree reasonably well. %
Again, this confirms that the Gaussian ensemble can reproduce multifractal statistics, also in the case of velocity gradients and their scaling with Reynolds number.

\section{Conclusion} \label{sec:6_conclusions}

The central contribution of this work is that it establishes a connection between multifractal models and functional approaches to turbulence based on ensembles of Gaussian fields. We developed an approach that promotes a multifractal model for increment statistics to a model for entire fields. This is accomplished by a change of variables that transforms a scale-dependent distribution of fluctuating H\"older exponents to a scale-independent distribution over the new parameter $\gamma$. This allows for the superposition of Gaussian characteristic functionals parameterized on $\gamma$, which then enables the computation of arbitrary field statistics, involving velocity, gradients or increments, for instance. %

In the current work, we first demonstrated an analytically-tractable instance of the transformation procedure that reproduces velocity increment statistics consistent with the lognormal model. After that we introduced a more general form of the transformation  that can encode multifractal increment and gradient statistics for any singularity spectrum $D(h)$. The resulting characteristic functional can be used to model data from direct numerical simulations (or experiments). 

Beyond the present work, the hope is that characteristic functionals as constructed here could also be helpful in the context of functional approaches to turbulence such as the Hopf equation \citep{hopf1952jrma}, which is an elegant way to encode the field statistics of turbulence.
A major open question towards this goal is the inclusion of skewness into the current framework, which is crucial to model the effects of self-advection and energy transfer across scales.
Another topic for future investigation is the prediction of multiscale correlations of velocity increments \citep{benzi1998multiscale,benzi1999multiscale,friedrich2018multiscale} by the characteristic functional models constructed in this work.

% % 
% 
% 

%

\backsection[Acknowledgements]{We thank Jan Friedrich for helping us to establish the connection of our transform to their work. Additionally, we thank Mostafa Kamal for his simulation data used for the velocity gradient scaling exponent comparisons.}

\backsection[Funding]{ 
This project has received funding from the European Research Council (ERC) under the European Union's Horizon 2020 research and innovation programme (Grant agreement No.\ 101001081).
}

\appendix
\section{Connection to the ensemble by Friedrich et al. (2021)}\label{app:Friedrich}
In their work on the explicit construction of joint multipoint statistics, \citet{friedrich2021jpc} constructed an ensemble of characteristic functionals exhibiting statistics of the Kolmogorov-Obukhov model~(K62; \citeauthor{obukhov1962some} \citeyear{obukhov1962some}; \citeauthor{kolmogorov1962refinement} \citeyear{kolmogorov1962refinement}). The K62 model is formulated in terms of $\varepsilon_\ell$, the energy dissipation rate coarse-grained over scale $\ell$. For simplicity, we focus on the inertial range, $\eta \ll \ell \ll L$, with $\eta$ the classic Kolmogorov dissipative scale, independent of $h$. Increment statistics in K62 can be implemented as a superposition of Gaussian distributions~\citep{castaing1990velocity,friedrich2025habil},
\begin{align}
    f_v(v; \ell) &= \int \mathrm{d}\varepsilon_\ell\, f_\varepsilon(\varepsilon_\ell; \ell) g\left(v; S(\ell; \varepsilon_\ell)\right)\,,
\end{align}
with
\begin{align}
    S(\ell; \varepsilon_\ell) &= U^2 \left(\frac{\varepsilon_\ell}{\overline{\varepsilon}}\right)^{2/3} \left(\frac{\ell}{L}\right)^{2/3}\,.
\end{align}
The distribution of $\varepsilon_\ell$ is assumed to be a scale-dependent lognormal distribution
\begin{align} \label{eq:epsell_pdf}
    f_\varepsilon(\varepsilon_\ell; \ell) = \frac{1}{\varepsilon_\ell \sqrt{2\pi \sigma^2(\ell)}} \exp\left(-\frac{(\log(\varepsilon_\ell) - M(\ell))^2}{2\sigma^2(\ell)}\right)
\end{align}
with location parameter
\begin{align}
    M(\ell) = \log(\overline{\varepsilon}) - \frac{\sigma^2(\ell)}{2}
\end{align}
and scale parameter 
\begin{align}
    \sigma^2(\ell) = \mu \log\left(\frac{L}{\ell}\right)\,.
\end{align}
The K62 model is equivalent to a multifractal model with quadratic singularity spectrum~\eqref{eq:quadratic_d_of_h} and without bounds on $h$. The difference is that $\varepsilon_\ell$ is used as a fluctuating parameter instead of $h$. The equivalence can be seen by applying the transformation $\varepsilon_\ell = \overline{\varepsilon} \left(\ell/L\right)^{3h-1}$ to~\eqref{eq:chevillard_second_order_sf}, \eqref{eq:incr_pdf_h_superposition}, and \eqref{eq:h_pdf_general} and setting $h_0 = 1/3 + \mu/6$ and $\sigma_h^2 = \mu/9$ while taking the inertial range limit.

With this in mind, we can apply the methodology outlined in \cref{sec:3_general_transform} to the K62 model, while replacing each instance of $h$ by $\varepsilon_\ell$. Contrary to \cref{sec:4_lognormal}, we choose a unit lognormal distribution for $\gamma$, whose CDF is
\begin{align}
    F_\gamma(\gamma) = G(\log(\gamma))\,.
\end{align}
The CDF of the fluctuating parameter $\varepsilon_\ell$ with scale-dependent PDF~\eqref{eq:epsell_pdf} reads
\begin{align}
    F_{\varepsilon_\ell}(\varepsilon_\ell; \ell) = G\left(\frac{\log(\varepsilon_\ell) - M(\ell)}{\sigma(\ell)}\right)\,.
\end{align}
Then, the structure functions of the $\gamma$-ensemble can be computed by~\cref{eq:s_tilde_general_transform} as
\begin{align}
    \widetilde{S}(\ell; \gamma) &= S\left(\ell; F_{\varepsilon_\ell}^{-1}(F_\gamma(\gamma); \ell)\right) \\
    &= U^2 \left(\frac{L}{\ell}\right)^{-2/3-\mu/3} \gamma^{\frac23 \sqrt{\mu \log\left(\frac{L}{\ell}\right)}} \label{eq:friedrich_subens_struc}
\end{align}
This turns out to be the same inertial-range form of the structure function as proposed in \citet{friedrich2021jpc}. As the basis for their model, they use fractional Ornstein-Uhlenbeck processes with H\"older exponent $H=1/3$. These are Gaussian processes with a specific, analytically known correlation function. For our considerations, it is only important to know that the corresponding structure function scales as
\begin{align} \label{eq:k41_strucfunc}
    S_\mathrm{K41}(\ell) \sim \ell^{2H} = \ell^{2/3}
\end{align}
for $\ell \ll L$, which is K41 scaling~\citep{kolmogorov1941a}. Then, the subensemble structure functions~\eqref{eq:friedrich_subens_struc}, which we obtained by application of our method to the K62 model, can also be obtained by applying the following length-scale transform to the K41 structure functions~\eqref{eq:k41_strucfunc}:
\begin{align}
    \ell \rightarrow \gamma^{\sqrt{\mu \log\left(\frac{L}{\ell }\right)}} \left(\frac{L}{\ell}\right)^{-\mu/2} \ell\,.
\end{align}
This is very close to the transform proposed in~\citet[eq. (10)]{friedrich2021jpc} for a fluctuating parameter $\xi \equiv \gamma$.
Overall, this means that the length scale transform proposed by~\citet{friedrich2021jpc} can be considered a special case of the more general transform presented here. However, choices regarding the small- and large-scale cut-off were made differently.

\section{Generalization to vector fields}
\label{app:vector_fields}

Our ensemble approach can be straightforwardly generalized to vector fields. To illustrate that, consider a three-dimensional incompressible velocity field $\bm u(\bm x)$. Analogous to \cref{eq:characteristic_functional}
the characteristic functional generalizes to
\begin{equation}
    \label{eq:characteristic_functional_vector field}
    \Phi[\bm \alpha; \gamma] = \left\langle \exp\left[ \imagi \int_{-\infty}^{\infty} dx \, \alpha_i(x) u_i(x) \right] \right\rangle
\end{equation}
where summation over repeated indices is implied and $\bm \alpha(\bm x)$ is a three-dimensional test function.  If $\bm u(\bm x)$ is an incompressible statistically homogeneous Gaussian vector field with zero mean, its characteristic functional takes the form
\begin{equation}
\label{eq:gaussian_characteristic_functional_vector_field}
    \Phi^{G}[\bm \alpha] = \exp \left[ -\frac{1}{2}\int_{-\infty}^{\infty} \dd \bm x\int_{-\infty}^{\infty} \dd \bm x' \alpha_i(\bm x) \widetilde C_{ij}(\bm x-\bm x';\gamma) \alpha_j(\bm x') \right],
\end{equation}
analogous to \cref{eq:gaussian_characteristic_functional}. Assuming statistical isotropy and incompressibility, the covariance tensor $\widetilde C_{ij}(\bm x-\bm x';\gamma) = \langle u_i(\bm x) u_j(\bm x') \rangle$ can be written as~\citep{pope2000turbulence}
\begin{equation}
  \widetilde{C}_{ij}(\bm r;\gamma) = %
  \sigma_u^2
  \left( \left[ \tilde f(r; \gamma ) + \frac{1}{2} r \tilde f'(r;\gamma) \right ]\delta_{ij} - \frac{1}{2} r \tilde f'(r;\gamma) \, \hat r_i \hat r_j \right)
\end{equation}
where $\tilde f(r;\gamma)$ is the longitudinal velocity autocorrelation function, prime indicates a derivative with respect to $r$; $\hat{\bm r}$ denotes the unit vector in the direction of $\bm r$.

As a next step, we can relate the longitudinal autocorrelation function to the conditional structure function
\begin{equation}
  \tilde f(r;\gamma) = 1 - \frac{1}{2 \sigma_u^2} \tilde S(r;\gamma)
\end{equation}
which then allows to establish the relationship between a multifractal model for the increment PDFs as outlined in \cref{sec:3_general_transform}. Once the covariance tensor is determined this way, the Gaussian ensemble \cref{eq:Gaussian_fields_ensemble} straightforwardly generalizes.

Quite obviously, addressing the question of how to include skewness into this framework is even more relevant for the generalization to vectorial statistics. This is, however, beyond the scope of the current work.

\bibliographystyle{jfm}
\bibliography{jfm}

\end{document}